\documentclass[a4paper,fleqn]{cas-dc}
\usepackage[numbers]{natbib}
\usepackage{natbib}
\setcitestyle{open={[},close={]}}
\usepackage[left]{lineno}
\usepackage{balance}

\def\tsc#1{\csdef{#1}{\textsc{\lowercase{#1}}\xspace}}
\tsc{WGM}
\tsc{QE}
\tsc{EP}
\tsc{PMS}
\tsc{BEC}
\tsc{DE}

\begin{document}
\let\WriteBookmarks\relax
\def\floatpagepagefraction{1}
\def\textpagefraction{.001}
\shorttitle{EEG Motion Detection}
\shortauthors{Adhikary et~al.}

\title [mode = title]{Optimized EEG Based Mood Detection with Signal Processing and Deep Neural Networks for Brain-Computer Interface}                      
\author[1]{Subhrangshu Adhikary}[orcid=https://orcid.org/0000-0003-1779-3070]
\cormark[1]
\address[1]{$^{,*}$Department of Research \& Development, Spiraldevs Automation Industries Pvt. Ltd., Raignaj, Uttar Dinajpur, West Bengal, India -733123 \\
subhrangshu.adhikary@spiraldevs.com}

\author[2]{Kushal Jain}
\address[2]{Resident Doctor, Vardhman Mahaveer Medical College and Safdarjung Hospital, New Delhi, India - 110029 \\
dr.kushalofficial@gmail.com}

\author[3]{Biswajit Saha}
\address[3]{Department of Computer Science and Engineering, Dr. B.C. Roy Engineering College, Durgapur, West Bengal, India -713206 \\
biswajit.saha@bcrec.ac.in}

\author[4]{Deepraj Chowdhury}
\address[4]{Department of Electronics and Communication Engineering, International Institute of Information Technology Naya Raipur, Naya Raipur \\
deepraj19101@iiitnr.edu.in}

\cortext[cor1]{Corresponding author: Subhrangshu Adhikary 
}

\begin{abstract}
Electroencephalogram (EEG) is a very promising and widely implemented procedure to study brain signals and activities by amplifying and measuring the post-synaptical potential arising from electrical impulses produced by neurons and detected by specialized electrodes attached to specific points in the scalp. It can be studied for detecting brain abnormalities, headaches, and other conditions. However, there are limited studies performed to establish a smart decision-making model to identify EEG's relation with the mood of the subject. In this experiment, EEG signals of 28 healthy human subjects have been observed with consent and attempts have been made to study and recognise moods. Savitzky-Golay band-pass filtering and Independent Component Analysis have been used for data filtration.Different neural network algorithms have been implemented to analyze and classify the EEG data based on the mood of the subject. The model is further optimised by the usage of Blackman window-based Fourier Transformation and extracting the most significant frequencies for each electrode. Using these techniques, up to 96.01\% detection accuracy has been obtained.
\end{abstract}

\begin{graphicalabstract}
\includegraphics{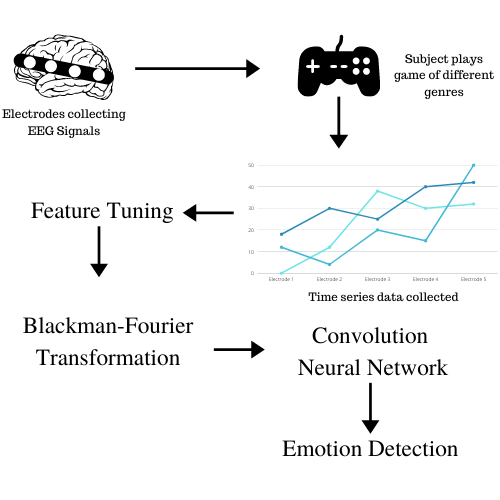}
\end{graphicalabstract}


\maketitle
\section{Introduction}
\label{sec:intro}
\begin{figure}
    \centering
    \includegraphics[scale=0.35]{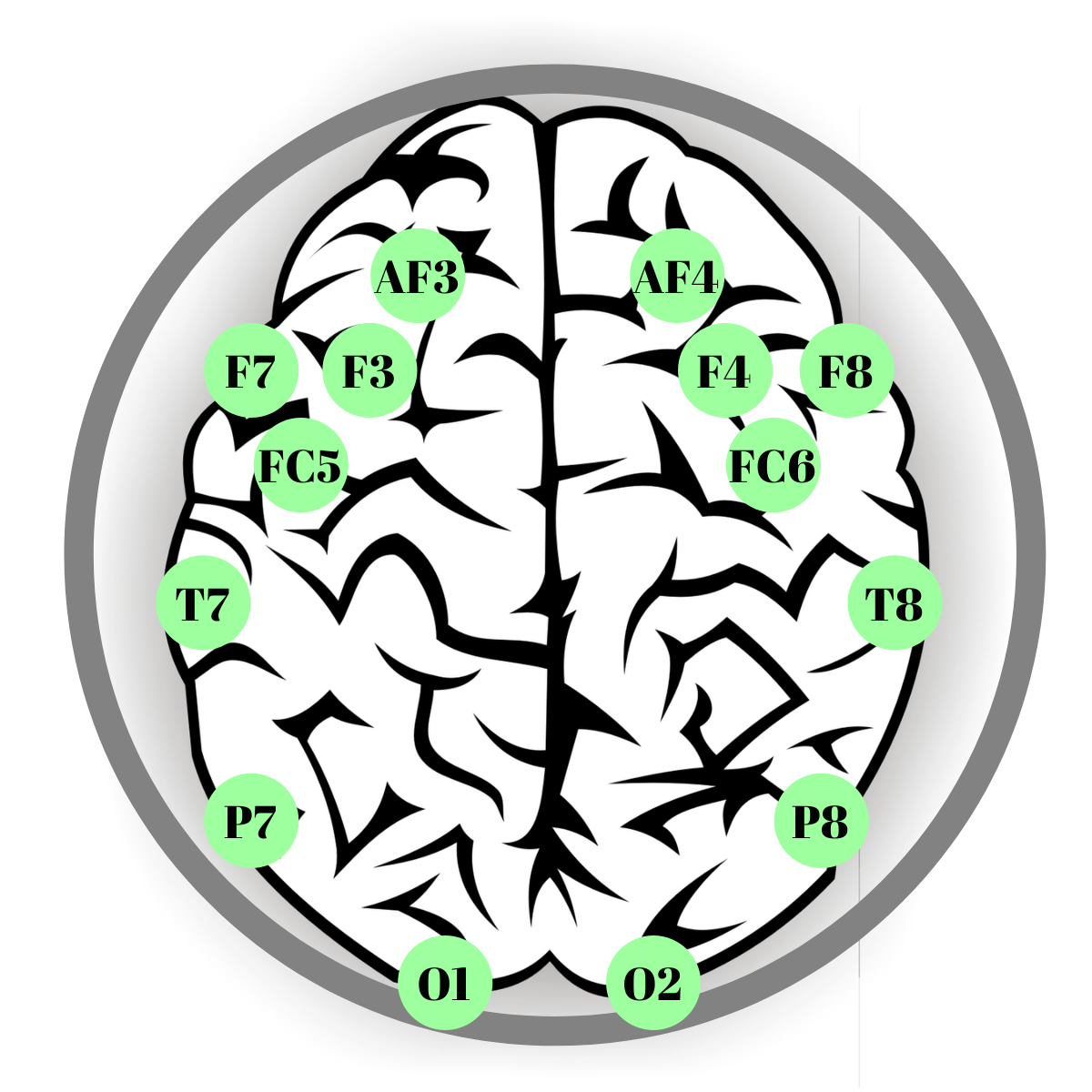}
    \caption{EEG electrodes placements for 14 electrodes used for the experiment and the region in the brain the placement corresponds to and named according to MCN nomenclature}
    \label{fig:eeg-mcn-placement}
\end{figure}

Human brains consist of chains of interconnected neurons which transmits messages by passing electrical signals. Due to different circumstances, like hyperactive neurons, inflammation, tissue damage, etc. neurons can pass abnormal impulses ~\cite{LEE20181490}. Electroencephalogram is a method by which specialized electrodes are connected to different parts of the brain to amplify and record the magnitude of the post-synaptic potential. Different frequencies of the electrodes decipher impulses of different neuron clusters of the brain. The recorded signals can be then visualized to extract several features helpful in understanding brain waves~\cite{LUO2019103132}. International measures have been implemented to standardize the electrode placement points and their identification known as Modified Combinatorial Nomenclature (MCN). Electrodes placed at Pre-Frontal, Frontal, Temporal, Parietal, Occipital and Central lobes are denoted by "Fp", "F", "T", "P", "O" and "C" respectively. The naming scheme of intermediate points of two consecutive major lobes has additional rules. "AF" series lies from "Fp" to "F", "FC" between "F" and "C", "FT" between "F" and "T", "CP" between "C" and "P", "TP" between "T" and "P" and finally "PO" between "P" and "O". Fig. \ref{fig:eeg-mcn-placement} shows the placement position of the electrodes with naming according to MCN nomenclature that has been used for the experiment ~\cite{faber2018two,9204431}.

\begin{figure*}
    \centering
    \includegraphics[scale=0.6]{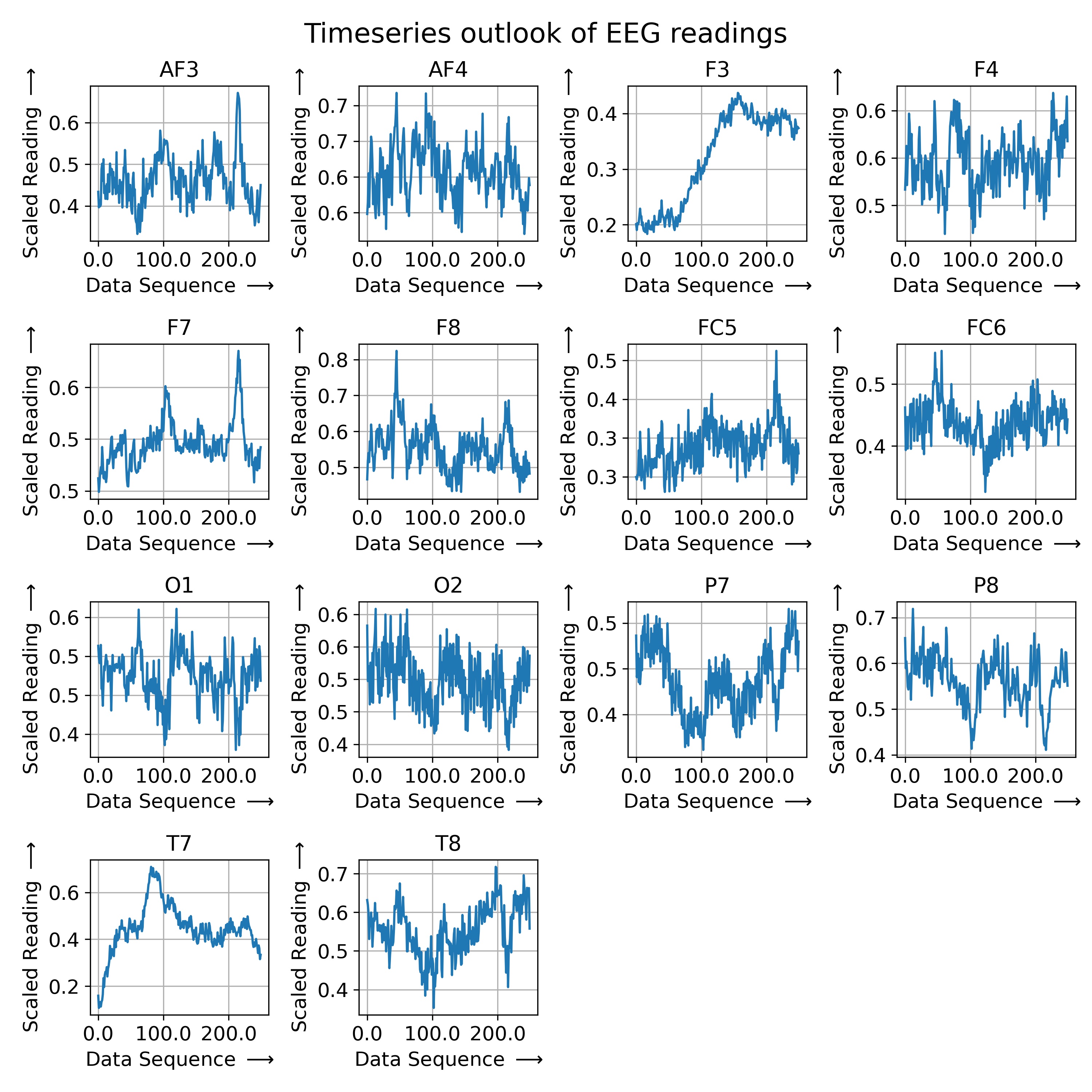}
    \caption{Sample of EEG recordings for the 14 electrodes placed with MCN nomenclature where the x axis represents the scaled magnitude of the reading and y axis represents the i'th element of the sequence where each element corresponds to 0.0078125 seconds}
    \label{fig:closeup}
\end{figure*}

As the brain controls our thinking and moods, therefore studying the brain signals can also indicate the mood of an individual. Several studies have been performed earlier to understand the brain signal to understand mood, however, there are very few approaches to establishing a smart decision-making system to understand the mood of a human~\cite{9609504}. Therefore with this experiment, attempts have been made to create an artificially intelligent decision-making system to classify the mood of a subject by means of the Neural Network algorithm. This is beneficial in many ways to understand consumer behaviour, and implement in road safety. For this purpose, a dataset has been obtained for 28 subjects on whom the experiment has been performed with their consent. There they are asked to play 4 different games while they were being monitored with EEG. After each game, they were asked about their present state of mind and their responses have been analyzed with the EEG recorded data in this experiment by implementing a smart decision-making classifier based on a neural network algorithm. Fig. ~\ref{fig:closeup} shows a sample of the EEG recordings for the 14 electrodes as discussed earlier. The x-axis represents the sequence of data and the y-axis represents the min-max scaler normalized readings. The device had recorded the data at a sampling rate of 2048 Hz which have been down sampled to 128Hz. Hence each element in x-axis represents 0.0078125 seconds of time.

Artificial Intelligence has evolved lately with the advancements in computing power. Popular AI technique includes Machine Learning. However classical machine learning algorithms like K-Nearest Neighbour, and Support Vector Machine performs doesn't perform optimally for large datasets because of their higher order time complexities. Therefore a specialized form of machine learning called a Neural Network can be implemented for processing data at a very large scale with almost linear time complexity~\cite{chakraborty_adhikary_ghosh_paul_n}. A neural network is inspired by the decision-making process of the human brain where data from one neuron is fed to other neurons by altering the electrical potential based on the properties of that neuron. Similarly, in artificial neuron, input in form of a matrix or tensor is fed to a neuron where it is multiplied with weights and added with biases resulting in different output matrix which is then passed through a chain of neural layers till a suitable output is found ~\cite{ACHARYA2018103,s20072034}. The ultimate goal of training is to reduce the difference between actual output and expected output known as cost function by different optimization algorithms like Stochastic Gradient Descent, Adam, RMSProp, etc. The training takes a very large time to find global minima of cost, but as the network size is fixed, therefore output generation with neural network is blazing fast. Convolution Neural Network (CNN) is a special form of neural network that specializes in image data and works by scanning the image to extract several distinguishable features with several different image processing techniques. Although this specializes in images, statistical data can be represented as a matrix as well and therefore CNN can work well with statistical data as well~\cite{9153955}.


Biomedical signal processing techniques play an important role in improving the reliability of the decision-making model. This is because the signal processing techniques are used to extract features from the signal that are unique to that class and this property helps in making the model robust ~\cite{ADHIKARY2022103321,suhaimi2020eeg}. Fourier transformation is a popular signal processing technique used to identify the frequencies that the signal is formed from. This would help to isolate the frequency ranges that are unique to the mood based on the EEG signals~\cite{ADHIKARY2022100277,10.1007/978-3-030-79065-3_30}. However, for more complicated data like EEG signals which are made up of several different frequency components and have an irregular pattern, it is difficult to identify the most significant frequency component of the signal by Fourier transformation. For this reason, another popular signal processing technique called a Blackman window can be incorporated with Fourier transformed EEG signal to amplify the most significant frequency components. The technique multiplies the Fourier-transformed dataset with a taper-shaped signal to minify non-significant frequencies ~\cite{9174858}. This would help to extract only the most significant frequencies for each EEG signal and therefore we can use this extracted data to train the model to make the model more robust and reliable compared to classifiers using raw data.

Based on these techniques, a smart and reliable Brain-Computer Interface could be made based on EEG signals and making the model robust by using deep learning models with biomedical signal processing techniques like Fast Fourier Transformation and further amplifying the results with the Blackman window. The primary contribution of the work includes:
\begin{itemize}
    \item To study the variation of EEG signals for different moods.
    \item To study the effect of the Blackman window and Fourier transformation on the EEG signals.
    \item To implement a neural network framework on the transformed data and compare the performance differences.
\end{itemize}

\section{Literature Survey}
\label{sec:backgroud}

\begin{table*}[h]
    \centering
    \begin{tabular}{p{1cm}|p{4cm}|p{2cm}|p{2cm}|p{4cm}}
        \hline
        Source & Objective & Algorithm & Performance & Limitation  \\
        \hline
        Acharya et. Al. ~\cite{10.1007/978-981-16-0401-0_38} & Multiple mood detection with Fast Fourier Transformation & LSTM \& CNN & 88\% accuracy & Training and testing dataset cross-contaminated with data from same sample space \\
        \hline
        Yin et. Al. ~\cite{YIN2021106954} & Mood recognition with fusion model of Graph based neural network & CNN \& LSTM & 90\% Accuracy & Binary Classification \\
        \hline
        Cui et. Al. ~\cite{CUI2020106243} & Mood detection by proposing Asymmetric Differential Layer (ADL) for feature extraction & RACNN & 95\% Accuracy & Binary classification \\
        \hline
        Naser et. Al. ~\cite{NASER2021102251} & Mood recognition based on music influence & \textit{k}-fold Cross-Validation & \textit{p}-value<0.05 & Trinary Classification \\
        \hline
        Liu et. Al. ~\cite{LIU20211} & Mood recognition with multi-channel textual and EEG features & SVM & Around 85\% Accuracy & Limited scalability for larger datasets due to exponential time complexity of SVM \\
        \hline
        Wang et. Al. ~\cite{WANG2021107626} & Mood recognition with Symmetric Positive Definite Matrix based feature extraction & SPDnet & p<0.001 & Trinary Classification \\
        \hline
        Tuncer et. Al. ~\cite{TUNCER2021110671} & Mood recognition with fractal pattern feature generation function & LDA, KNN \& SVM & 99.82\% Accuracy & Limited scalability because of exponential time complexity of the classifiers \\
        \hline
    \end{tabular} 
    \caption{Summary of the State of the Art showing relevant recent works for the subject area, their objectives, algorithms used, primary findings and limitations that have been addressed with the proposed work}
    \label{tab:comparison-table}
\end{table*}
EEG is a very popular technique for understanding brain activities. EEG has been earlier used to study the effects of damage on the pre-frontal cortex of the brain and its effects on mood understanding ~\cite{10.1093/brain/awx031}. Patients suffering from depressive disorders have shown asymmetry in EEG of the left and right pre-frontal cortex ~\cite{HAGHIGHI2017137}. EEG recordings from the pre-frontal cortex have also revealed that it can facilitate affective processing in absence of visual awareness ~\cite{10.3389/fnhum.2018.00412}. The frontal cortex has been mostly studied with EEG to detect epileptic seizures and this is very challenging as neocortical seizures spread rapidly ~\cite{FEYISSA2017157}. As the frontal cortex is responsible for voluntary muscle movement, it can be studied with EEG to understand multiple emotional clues which are related to voluntary muscle movements triggered both consciously or unconsciously ~\cite{byzova2020monitoring}. The temporal is responsible for auditory processing and implementing machine learning can detect microstate alteration in patients living with temporal lobe epilepsy ~\cite{V20188}. An experiment conducted to understand the latency of auditory stimulation and its effects on EEG readings has revealed latency of 41-52 ms for near 40Hz frequency and around 21-27 ms for frequency higher than 80Hz ~\cite{wang2021estimating}. The parietal lobe facilitates sensation and is a major indicator of mood ~\cite{li2019eeg}. Studies revealed that positive and negative moods are separable based on the EEG readings on this lobe ~\cite{10.1007/978-3-319-70093-9_90}. The occipital lobe helps to visualize, estimate depth and colour, recognize objects, etc. and this is an important area of the brain that can influence mood ~\cite{s18030841}. EEG of the occipital lobe has shown elevated activities on showing food products to the subjects triggering a mood elevation ~\cite{songsamoe2019understanding}.

Artificially intelligent decision-making has been implemented often by means of machine learning or deep learning neural networks. Machine learning has successfully classified Alzheimer’s disease based on EEG data ~\cite{ieracitano2020novel}. One-minute EEG data have also accurately detected Schizophrenia utilizing machine learning ~\cite{buettner2020development}. Studies have shown that age can be estimated with EEG recordings based on machine learning predictions ~\cite{10.3389/fnagi.2018.00184}. Quasi Recurrent Neural Network and Gated Recurrent Unit algorithms have been proven to outperform classical machine learning techniques in terms of accuracy and stability based on EEG gait sequences ~\cite{nakagome2020empirical}. Fast Fourier transformation have earlier been used with EEG signals to detect multiple moods with around 88\% accuracy. However, in the experiment, both the training and testing dataset contained signals from the same sample space making it lesser reliable in the uncontrolled environment ~\cite{10.1007/978-981-16-0401-0_38}. Experiments have been performed using neural networks on the EEG spectrum for identifying different features but they have not shown any robust signal processing method \cite{9204431,suhaimi2020eeg}. Transfer learning is another method of using deep learning in which the learned weights and biases are transferred to a new model that is used for faster convergence. Now, the state-of-the-art method with this process has limited classification accuracy that can be further improved ~\cite{s20072034}.

\subsection{Motivation for the experiment}

After a thorough investigation of the literature, it has been found that different works have been conducted on EEG for the identification of several properties ~\cite{DADEBAYEV2021}. It has been used to detect the fatigue level of humans using a smart decision-making framework. Biometric authentication systems have been made using EEG ~\cite{9609504}. EEG signals have been attempted to convert thoughts into text. Some works have also been performed to detect moods based on machine learning framework on raw EEG signals and other works have shown usage of different signal processing techniques with a threshold-based decision, making them lesser reliable for external datasets ~\cite{pandey2021subject}. Table ~\ref{tab:comparison-table} further summarizes the most relevant recent works related to the subject area and discusses their objective, performances and limitations that have further scopes for improvement. Therefore, this motivated to create a robust and reliable mood detection system based on EEG signals making use of the Blackman window on Fourier transformation and neural network framework.

\section{Methodology}
\label{sec:methods}
\begin{figure}
    \centering
    \includegraphics[scale=0.3]{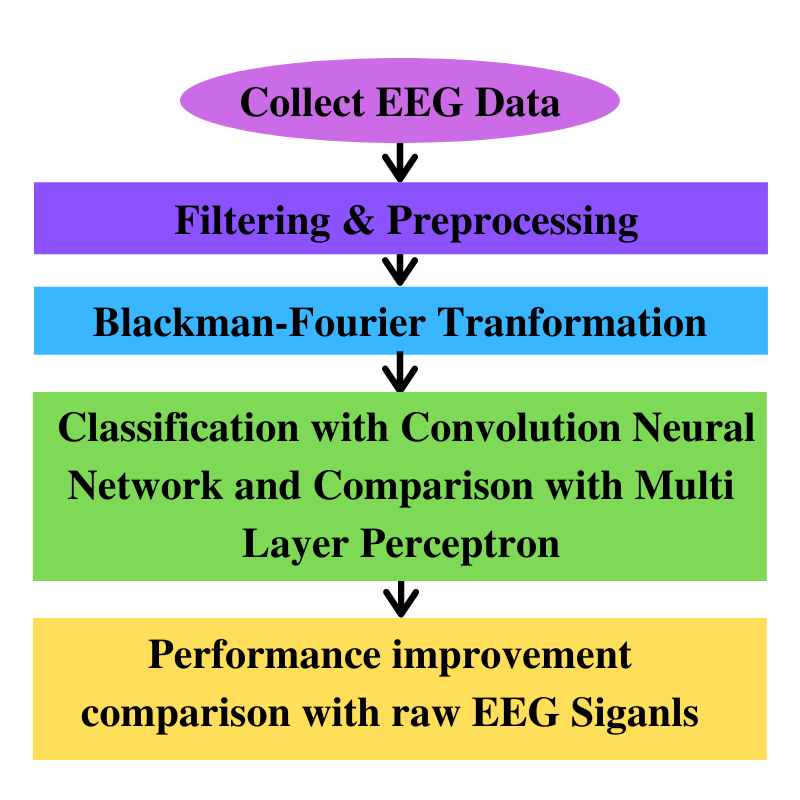}
    \caption{The workflow of the experiment}
    \label{fig:workflow}
\end{figure}

The workflow of the experiment consists of data collection, filtering \& preprocessing, transformation, classification and performance testing of the work. Fig. ~\ref{fig:workflow} shows the workflow of the experiment.

\subsection{Data Collection, Filtering \& Preprocessing}
The experiment has been conducted on an openly available dataset published by Elsevier under Creative Commons License \cite{ALAKUS2020101951}. The dataset consists of 28 subjects whose EEG data have been recorded with consent using a portable EEG device named 14 channel Emotiv Epoc+. The sampling rate of the device is 2048 Hz and the data have been down sampled to 128 HZ. The participants were from the Firat University Faculty of Technology in the Department of Software Engineering. They belong to the age group of 20-27 and they had no previous history of any psychological illness. The number of male or female candidates is confidential in the dataset and therefore we consider this study as general. They were asked to play 4 different video games of boring, calm, horror and funny genres while monitoring their brain activity with EEG for 20 minutes and at the end of the game, they were asked about their mood whether they liked the game, their satisfaction level on 1-10 scale, whether they had played the game earlier, do they have any memory of any event related with the game, etc. The experiment was conducted in a dark quiet room of the university and the game was shown on a notebook with a screen size of 15.6 inches. The EEG device collected data from 14 specific points based on MCN nomenclature namely AF3, AF4, F3, F4, F7, F8, FC5, FC6, O1, O2, P7, P8, T7, T8. The dataset contains a total of 4,284,224 rows and 14 columns which have been split into a training set having data from 54\% subjects, a validation set having data from 18\% subjects and a testing set having data from 28\% subjects. Based on their answers about satisfaction level, the 1-5 range was labelled as Low and the 6-10 was labelled as High. Similarly boring, horrible, calm and funny were labelled as High or Low based on the ratings provided by the subjects. A total of 4284224x14 data points have been considered for the experiment. This is to note that none of the three sets has data from common subjects. Further, the data have been passed through Savitzky-Golay (Savgol) filter and some portion of the data have been passed through Independent Component Analysis (ICA) before Blackman-Fourier processing. These have been explained below.

\subsubsection{Savitzky–Golay filtering}
The electrodes of the EEG devices are highly sensitive and can sense within the microvolt range. Because of this high sensitivity, the recordings from the EEG device are very susceptible to noises. Because of this, either Savitzsky-Golay band-pass filtering or Discrete Wavelet Transform is used to eliminate such noises. The presented experiment used Savitzsky-Golay filtering of order 3rd-degree because of its smoothing properties without distorting the signal tendency~\cite{9272776,s20030807}.

\subsubsection{Independent Component Analysis}
On the other hand, as different portions of the brain are responsible for different activities and several brain activities occur simultaneously, the electrodes of EEG devices often capture overlapped signals of multiple brain activities. Few placements are more susceptible to capturing brain activity caused by voluntary muscle movements, external stimuli, etc. with high intensity. This can cause extreme asymmetry in the results. Therefore this can be filtered out by independent component analysis. The presented experiment has particularly used ICA on F7 and T7 as these have the highest asymmetry in the distribution~\cite{STONE200259,https://doi.org/10.1049/iet-spr.2020.0025}.

\subsection{Games and their impact on mood}
Boring game in the study was chosen as Train Sim World. This game was considered as per global scores, comments and criticism. The game was designed in such a way that only those who know the details of the game can play the game and the participants didn't know how to play the game. The game was not understood by the participants and they randomly pressed different buttons. Low Arousal Negative Valence emotion was obtained from this method.

Calm game in the study was a game called Unravel. This was considered calm based on the kinds of music, sounds and the slow and relieving atmosphere of the game. The participants in the experiments were observed to understand the game mechanics and also listened to the music which stimulated their brains to reach a relaxed state. Low Arousal Positive Valence emotion was obtained from this method.

Horror game in the study was chosen as Slender – The Arrival. The atmosphere in the game was dark and quiet with very disruptive sounds. High Arousal Negative Valence emotion was obtained by this method.

Funny game in the experiment was called Goat Simulator. The atmosphere and sound of the game were very charming and most of the participants were observed to have a smile on their faces while playing the game. High Arousal Positive Valence emotion was extracted from this method.

\subsection{Defining the moods}
Traditionally either Discrete models or Valence-Arousal-Dominance (VAD) models are used for mood detection where valence describes the pleasantness of the stimuli, arousal determines the intensity of mood provoked by the stimuli and dominance describes the degree of control exerted by the stimuli. Although this model is suitable for many use cases, this increases the complexity of classifying the mood and therefore instead of using the VAD model, independent classes have been used each corresponding to a different mood. This enables the model to be built based on a categorical classification framework. This way, model complexity was reduced and enhanced the classification performance. The 5 classes of moods used in the model are Satisfaction, Boring, Horrible, Calm and Funny. The satisfaction levels mean the mood when the players obtained gaming performance meeting their expectations. Boring mood denotes the mood where the player finds the game tedious and repetitive. Horrible mood denotes the mental condition of the player where the player gets frightened because of the gameplay and their senses are elevated in preparing for a fight-or-flight scenario. Calm is the mood that arises while playing games where the players are not terrified or not performing any tedious repetitive work and the gaming performance depends much on reflexes rather than developing strategies. Finally, funny is a mood in which laughter is expected from the players.


\subsection{Blackman-Fourier Data Processing}
The datasets have been split into smaller chunks of \textit{N} elements each, here 128. The Fourier transformation has been applied with 1D Discrete Fast Fourier Transformation (FFT) method. Therefore for dataset \textit{k} with \textit{N} elements, eqn. ~\ref{eqn:FFT} represents the primary transformation function ~\cite{zavaleta2020proposal}.

\begin{equation}
    \label{eqn:FFT}
    y(k) = \int^{N-1}_{n=0} e^{-2\pi j \frac{kn}{N}}x[n]
\end{equation}
Following this, the Blackman window of 128, i.e., \textit{N} elements is given by eqn. ~\ref{eqn:blackman}.
\begin{equation}
    \label{eqn:blackman}
    w(n) = 0.42 - \frac{1}{2}\cos{\frac{2\pi n}{M} + 0.08\cos{\frac{4\pi n}{N}}}, \text{where } n = 1,2,...,N
\end{equation}
From eqn. ~\ref{eqn:FFT} and eqn. ~\ref{eqn:blackman}, the Blackman-Fourier transformation \textit{g(k)} have been obtained as eqn. ~\ref{eqn:blackman-fourier}.
\begin{equation}
    \label{eqn:blackman-fourier}
    g(k) = \int^{N-1}_{n=0} e^{-2\pi j \frac{kn}{N}}x[n]\times w(n)
\end{equation}

The eqn. ~\ref{eqn:blackman-fourier} returns a complex set of results and therefore, FFT Frequency has been found with a sample spacing of \textit{c = 0.0078125} to further isolate the significant frequencies (here, magnitude over 0.35) as defined by \textit{s(n)} in eqn. ~\ref{eqn:fft-freq}.
\begin{equation}
    \label{eqn:fft-freq}
    s(n) = \frac{c\times g(n)}{N}, \text{where } n = 1,2,...,N
\end{equation}
Following this, all the points of Blackman windowed FFT having magnitude higher than 0.35 have been considered to create a new transformed dataset. This method is iterated over the entire raw dataset in slices of size 128 rows. This dataset has been used to build the neural network model.

\subsection{Neural Network and Performance Evaluation}
The simplest form of a neural network is Multilayer Perceptron (MLP). A node or artificial neuron is also known as perceptron and the combination of multiple layers of neurons combines to form an MLP network ~\cite{8651516}. Adding convolution filters in combination with max-pooling layers on top of an MLP network, a Convolution Neural Network can be built which can parse the data mimicking an image matrix to find several patterns such as texture, colour combination, edges, etc. and later pool out the most contrasting features and amplify them to detect the class. And therefore it is supposed to work much better than a general MLP ~\cite{9210487}. Traditional machine learning classifiers have not been considered due to their limitations to handle large volume of data.

The network built for this experiment includes a one-dimensional convolution layer consisting of 32 filters having kernel size of 1 having hyperbolic tangent as an activation function in the first layer, followed by a max pooling layer which pools out 1 feature at a time. After this, the tensor has been flattened and to prevent the model from overfitting, used a 20\% dropout rate. Following these, 3 hidden layers of 500 nodes each have been created. The final layer of the network contains 5 nodes each corresponding to a different mood. Rectified linear unit (ReLU) activation function has been used for the hidden layers and the Softmax function has been used on the output layer because the analysis is converted into a categorical classification problem. Adam is an upgrade to the Stochastic Gradient Descent algorithm which converges the training quicker and hence this weight update rule has been implemented. The categorical cross-entropy loss function has been used along with Adam optimizer for separately detecting the different moods. The model has been trained until it reached a global minimum which generally happened somewhere between 5,000 to 10,000 epochs for different training combinations. Finally, to validate the model, the test data have been used to generate results which are then compared with the expected class and parameterized the performance based on several metrics such as accuracy, precision, recall, F1 score, cohen kappa score, jaccard similarity score, training and testing time.

\section{Results}
\label{sec:results}
\begin{figure*}
    \centering
    \includegraphics[scale=0.5]{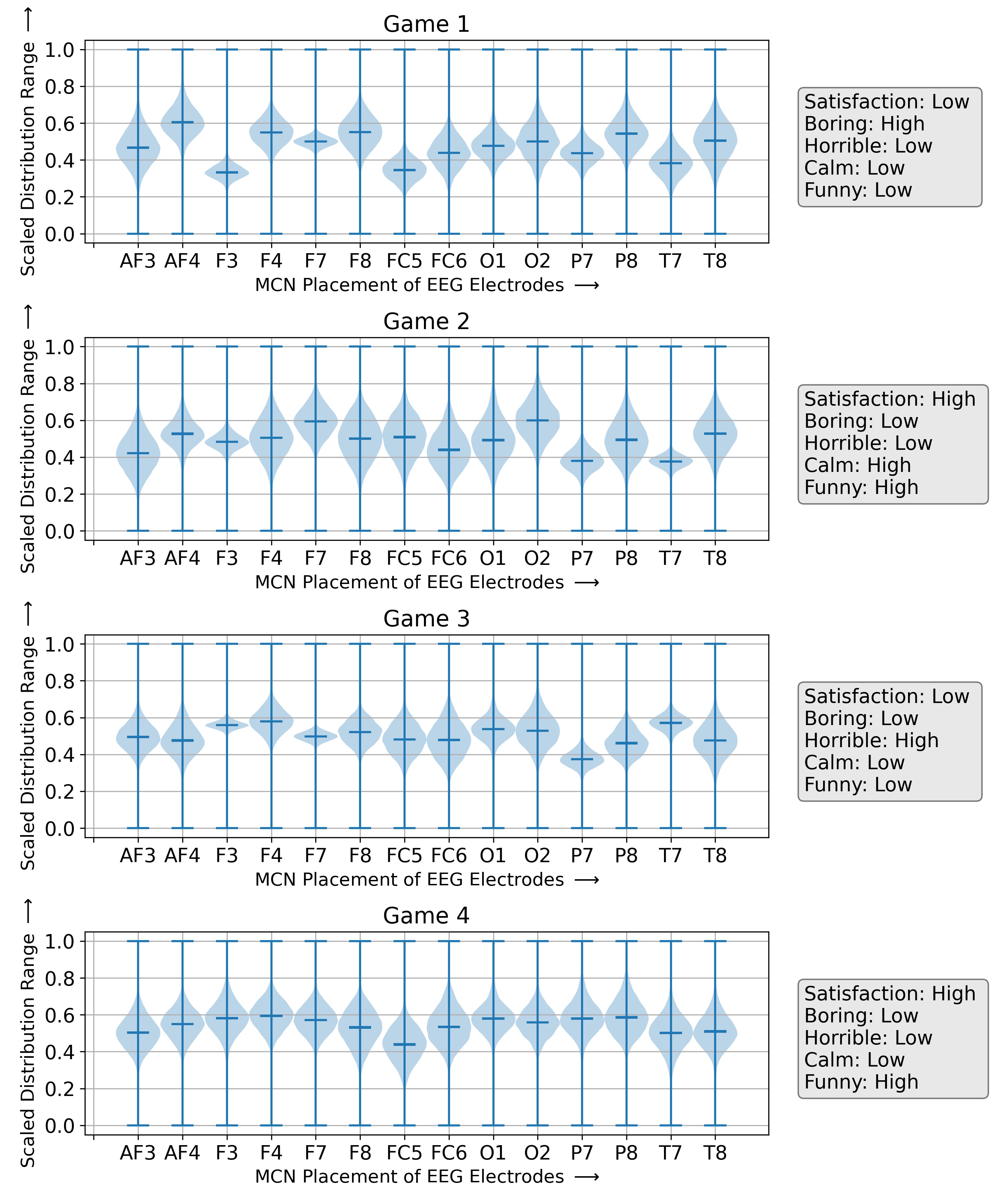}
    \caption{Violin distribution of the EEG recordings while playing 4 different games of various genres along with their ratings as described by the subject}
    \label{fig:violin-distribution}
\end{figure*}

The main aim of this experiment is to understand the classifiable properties of EEG data for various moods in order to establish a detection algorithm based on neural network techniques. For this purpose, after close observation of the distribution range fig. \ref{fig:violin-distribution} of a person while playing four different genres of games, it is apparent that the EEG distribution pattern is different. The distribution have been represented for one randomly picked subject while playing all different games. The data have been normalized with min-max scaler method which fits the entire dataset within the range of 0 and 1. The plots are made based on the data from each points of the recording for the subject for all different electrode placement. The bandwidth of F7 has the widest spectrum while the subject found the game funny. This appears from the frontal lobe of the brain which is mostly responsible for voluntary activities. This gives a clear indication that whenever the subject found the game funny, the activity of the frontal lobe increases inducing increased muscle activities. The F7 and T7 were found to have extreme asymmetry which indicates that these artefacts were contaminated by external stimuli or voluntary muscle movement, etc. For this purpose, ICA was applied on these to eliminate such discrepancies. This can also be noticed that the fun is related to the overall satisfaction with the subject. Studying the AF3 readings, it is noticed that whenever the subject played a horrible game, the AF3 spectrum got narrower indicating restricted activities in between the pre-frontal and frontal lobes. Whenever the subject played a boring game, the median of F3 readings fell below 0.35 mix-max scaler normalized units. The same happened with FC5 readings. These indicate that the frontal lobe has diminished activity resulting in much reduced and lethargic muscle movement. While playing a calm game, it has been observed that the O1 and O2 have the widest bandwidth indicating plenty of nerve activity involved with the occipital lobe. Many more such patterns could be identified from one or a combination of multiple electrode readings associated with similar kinds of moods which can be exploited for artificially intelligent detection of various moods in a smart environment.

\begin{figure*}
    \centering
    \includegraphics[scale=0.5]{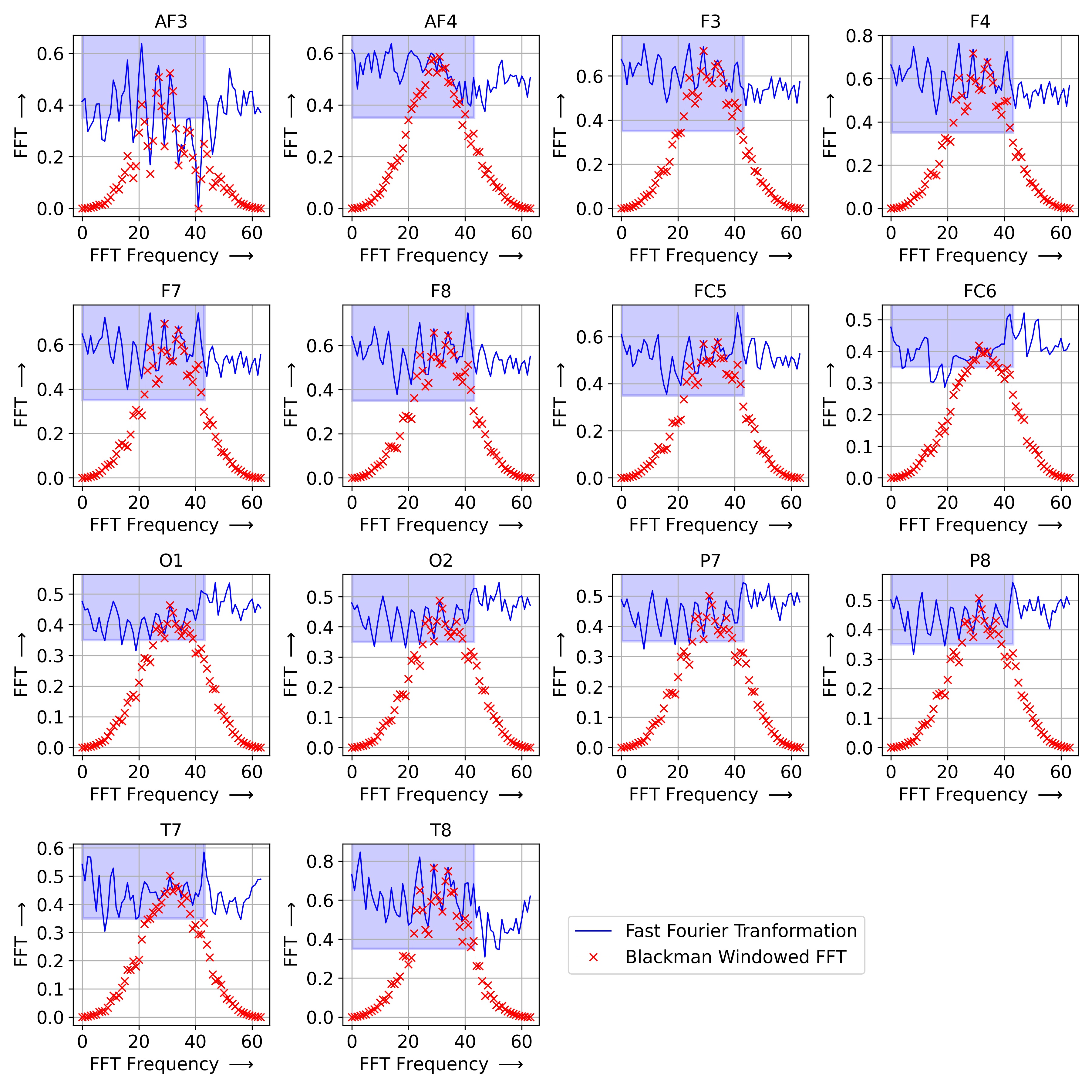}
    \caption{Representation of Fast Fourier Transformed data v/s Transformation frequency for both raw Fourier Transformation (represented with blue line) and Blackman windowed Tranformation (red line 'x' markers). Also, the blackman windowed FFT in highlighted region has been used for the classification.}
    \label{fig:fft}
\end{figure*}

The raw EEG data have been transformed by Fourier Transformation using the Blackman window after splitting the dataset with 128 elements each and one such chunk is shown in Fig. ~\ref{fig:fft}. Here the blue line shows the Raw Fourier Transformation using the Fast Fourier Transformation algorithm and the red line with 'x' marker shows the transformation with the Blackman window. The plot represents a sample of FFT and Blackman windowed FFT for a random subject while playing a random game. A slice of 128 rows simultaneously for all the placements has been picked at random from somewhere in the middle of the dataset. The data is then normalized with min-max scaler method and the corresponding alogrithms were applied and plotted. It can be noted that the FFT peaks of F7 are not clearly distinguishable, however, combined with the Blackman window, it can be found that the F7 signals have maximum dependency near a frequency of 30Hz. Similarly, all other electrodes can be transformed to find the most significant frequencies they correspond to. Following this, the most significant frequency bands for each electrode have been considered for the classification as this provided a robust and reliable detection. The Blackman-FFT points in the highlighted region in the graph have been considered for further classification.

\subsection{Classification Performance for Blackman-Fourier Transformation}
Neural networks are powerful classifiers for training a very large dataset. This works as a series of matrix multiplication and specialized hardware such as a Graphics Processing Unit or Tensor Processing Unit can drastically improve the training speed. To utilize this property in a robust and reliable manner, Blackman windowed FFT has been used. For this purpose, the experiment has been divided into a categorical classification problem for detecting High or Low labels of the 5 moods namely Satisfaction, Boring, Horrible, Calm and Funny. The data have been tuned to extract features as discussed in the earlier sub-section. After training, validation and testing, the performances of the convolution neural network have been recorded in table ~\ref{tab:best-parameter-performance}. From this, it is apparent that the algorithm has successfully detected boredom level with the highest accuracy of 96.01\% followed by moods such as calm, horrible, funny, and satisfaction with 95.41\%, 95.13\%, 94.85\% and 94.12\% accuracy. However, accuracy doesn't properly describe the false errors for an imbalanced dataset and for this purpose, let us look at other metrics as well. It can be observed that calm has the highest precision (0.934) indicating very low chances of marking other moods as calm. The boring mood has also a similar precision score (0.933). It can be noticed that although the accuracy of detecting horrible and funny experiences is higher than the accuracy of detecting satisfaction levels, the precision is satisfaction (0.912) is higher than horrible (0.904) and funny (0.909). This denotes the presence of a biased quantity of data for each mood that is more likely to be found in a real-world scenario. The recall is highest for detecting boring experiences (0.956) indicating a very lesser amount of leaving an actual boring experience than marking it as a boring experience. Recall of calm (0.947) and horrible (0.927) experience are high as well, funny is decent (0.922) but satisfaction is lowest of all (0.908). The recall order for the experiences follows the same order as for accuracy. F1 score is the harmonic mean of precision and recall and for detection of the experiences it follows the order Boring (0.944), Calm (0.940), Horrible (0.915), Funny (0.915) and Satisfaction (0.909). The Jaccard Similarity score shows the similarity or diversity with increasing similarity toward jaccard score 1 and increasing diversity toward jaccard score 0. The score follows the order same as the accuracy with Boring 0.899, Calm 0.872, Horrible 0.847, Funny 0.843 and Satisfaction 0.831. Cohen-Kappa score shows the inter-relatability of true and detected labels and for the experiment, this also follows the same order as accuracy with Boring 0.893, Calm 0.862, Horrible 0.831, Funny 0.826 and Satisfaction 0.812.

\begin{table*}[h]
    \centering
    \caption{Classification performance for mood detection with Blackman-Fourier transformation and Convolution Neural Network}
    \begin{tabular}{c|c|c|c|c|c}
        \hline
        Metric & Satisfaction & Boring & Horrible & Calm & Funny \\
        \hline
        Accuracy (\%)      & 94.12 & 96.01 & 95.13 & 95.41 & 94.85 \\
        Precision          & 0.912 & 0.933 & 0.904 & 0.934 & 0.909 \\
        Recall             & 0.908 & 0.956 & 0.927 & 0.947 & 0.922 \\
        F1 Score           & 0.909 & 0.944 & 0.915 & 0.940 & 0.915 \\
        Jaccard Similarity & 0.831 & 0.899 & 0.847 & 0.872 & 0.843 \\
        Cohen-Kappa        & 0.812 & 0.893 & 0.831 & 0.862 & 0.826 \\
        Train Time (s)     & 323.1 & 482.5 & 465.8 & 518.1 & 542.1 \\
        Testing Time (s)   & 0.252 & 0.272 & 0.225 & 0.302 & 0.226 \\
        \hline
    \end{tabular}
    \label{tab:best-parameter-performance}
\end{table*}

Apart from the detection performances, emphasis should also be given to the resource consumption for training and testing in order to plan the hardware setup and cost. It was noticed that training for Funny mood took the longest duration of 542.1s followed by Calm 518.1s, Boring 482.5s, Horrible 465.8s and Satisfaction 323.1s. This indicates a decreasing order complexity to reach the global minimum of loss. As for the detection, Horrible experience was detected the fastest with 0.225s detection time followed by Funny, Satisfaction, Boring and Calm by taking 0.226s, 0.252s, 0.272s and 0.302 respectively.

\begin{table*}[h]
    \centering
    \caption{Classification performance for mood detection with Blackman-Fourier transformation and Multi-Layer Perceptron}
    \begin{tabular}{c|c|c|c|c|c}
        \hline
        Metric & Satisfaction & Boring & Horrible & Calm & Funny \\
        \hline
        Accuracy (\%)      & 86.10 & 89.80 & 88.92 & 89.95 & 87.11 \\
        Precision          & 0.802 & 0.829 & 0.810 & 0.842 & 0.818 \\
        Recall             & 0.814 & 0.855 & 0.842 & 0.851 & 0.830 \\
        F1 Score           & 0.807 & 0.842 & 0.825 & 0.846 & 0.824 \\
        Jaccard Similarity & 0.672 & 0.726 & 0.708 & 0.670 & 0.641 \\
        Cohen-Kappa        & 0.595 & 0.666 & 0.639 & 0.670 & 0.641 \\
        Train Time (s)     & 477.7 & 558.6 & 557.0 & 629.6 & 611.2 \\
        Testing Time (s)   & 0.292 & 0.332 & 0.319 & 0.311 & 0.290 \\
        \hline
    \end{tabular}
    \label{tab:raw-data-higher}
\end{table*}
The transformed dataset has been compared with a regular MLP network as well to further validate the effectiveness of the transformation. Based on this aspect, MLP best detected Calm mood with 89.95\% accuracy followed by Boring with 89.80\% accuracy, Horrible mood with 88.92\% accuracy, Funny mood with 87.11\% accuracy and Satisfaction level with 86.10\% accuracy. Precision follows the order Calm (0.842), Boring (0.829), Funny (0.818), Horrible (0.810) and Satisfaction (0.802). The recall follows the order Boring (0.855), Calm (0.851), Horrible (0.842), Funny (0.830) and Satisfaction (0.814). Further looking at the training time, Calm mood took the longest to train as it takes 629.6s to build the model followed by Funny with 611.2s, Boring with 558.6s, Horrible 557.0s and Satisfaction with 477.7s. Finally, the testing time was the fastest for Funny mood taking 0.290s to predict results followed by Satisfaction with 0.292s, Horrible mood with 0.316, Calm mood with 0.319s and Boring mood with 0.332s.

\subsection{Performance comparison for raw data Classification}
To compare the performance upgrade for a regular convolution neural network on raw data, table ~\ref{tab:raw-performance} records the performance for mood detection from raw EEG data with CNN. It can be noticed that performances are quite poor compared to the proposed method. 83.78\% was the highest accuracy obtained and that is to determine the satisfaction level. Accuracy of other experiences follows in order Horrible, Boring, Calm and Funny having accuracy of 80.94\%, 79.74\%, 79.13\% and 78.68\%. Precision was highest for detecting horrible experience (0.716) and lowest for detecting funny mood (0.591). The recall was highest for satisfaction (0.827) and lowest for horrible (0.745). Likewise, the F1 score found was highest for Satisfaction and lowest for funny (0.597). Here it’s important to note that the training time for CNN with raw data was generally several times higher than the training and testing time for CNN on optimized data. This is due to the fact that the transformation has reduced the dataset sizes and training or testing on lesser data took faster epochs. This ultimately speeds up the model training for transformed data and therefore the performance on raw data is much slower as discussed. The training required the longest to converge for Funny mood taking 6187s to train, following this, Calm took 5945s to train, Horrible took 5545s, Boring took 5272s and Satisfaction took 4354s. Discussing the testing time, Horrible mood was detected the fastest with detection made within 2.04s, followed by Boring with 2.21s, Satisfaction with 2.31s, Calm with 2.44s and Funny with 2.56s.

Therefore, it can be clearly understood that the Blackman-Fourier transformation has significantly improved the model accuracy compared to categorical classification with raw data.

\begin{table*}[h]
    \centering
    \caption{Classification performance for mood detection with Raw Data and Convolution Neural Network}
    \begin{tabular}{c|c|c|c|c|c}
        \hline
        Metric & Satisfaction & Boring & Horrible & Calm & Funny \\
        \hline
        Accuracy (\%) & 83.78 & 79.74 & 80.94 & 79.13 & 78.68 \\
        Precision & 0.706 & 0.632 & 0.716 & 0.607 & 0.591 \\
        Recall & 0.827 & 0.759 & 0.745 & 0.780 & 0.781 \\
        F1 Score & 0.738 & 0.652 & 0.728 & 0.619 & 0.597 \\
        Jaccard Similarity & 0.611 & 0.526 & 0.594 & 0.500 & 0.483 \\
        Cohen-Kappa & 0.486 & 0.327 & 0.457 & 0.278 & 0.243 \\
        Train Time (s) & 4354 & 5272 & 5545 & 5945 & 6187 \\
        Testing Time (s) & 2.31 & 2.21 & 2.04 & 2.44 & 2.56 \\
        \hline
    \end{tabular}
    \label{tab:raw-performance}
\end{table*}

\section{Discussions}
The results obtained from the experiment show considerable improvement compared to the State of the Art. This section discusses the results comparison, quality control documentation, implementation scopes and practical usage of the work in real-world scenarios and finally the limitations.

\subsection{Comparison with State of the Art}
The proposed work has a considerable contribution toward the state of the art. Earlier, Tao et. Al. \cite{9204431} have shown the usage of Neural Network techniques on the EEG spectrum however, no specific feature tuning or signal processing technique has been used making it lesser reliable for a wider spectrum of data, our work has shown the usage of signal processing technique making it theoretically more reliable and robust when exposed to the uncontrolled environment. Cimtay et. Al. ~\cite{s20072034} have shown the usage of the transfer learning method for faster convergence of the model while recognizing mood, however, the work has obtained up to 86.56\% accuracy whereas the proposed work has obtained up to 96.01\% accuracy. A state-of-the-art review ~\cite{suhaimi2020eeg} that has been recently published shows that most of the similar works that have been performed have used classification methods on raw EEG signals without proper signal processing and therefore the proposed work has improved this study gap by introducing a more reliable framework. Further, the work done by Acharya et. Al. ~\cite{10.1007/978-981-16-0401-0_38} had obtained 88\% detection accuracy but the training and testing datasets have been contaminated by data with the same sample space. However, the presented model obtained 96.01\% accuracy while training and testing the model with data from a different group of people. The work of Yin et. Al. ~\cite{YIN2021106954} and Cui et. Al. ~\cite{CUI2020106243} had detected only two classes of moods and Naser et. Al. ~\cite{NASER2021102251} and Wang et. Al. ~\cite{WANG2021107626} had detected 3 classes of moods. However, the proposed model has successfully detected 5 classes of moods. The 
work presented by Liu et. Al. ~\cite{LIU20211} was classified with SVM and work presented by Tuncer et. Al. ~\cite{TUNCER2021110671} was classified by LDA, KNN and SVM. All of these machine learning algorithms have limited scalability because of their exponential training or prediction time complexity making it practically inappropriate for usage in a real-world scenario.

\subsection{Quality control documentation}
The work has been conducted based on Open Source Repositories which have earlier been cited by peer-reviewed works. The data collected by the device had a resolution of $0.1275\pm1 \mu V$ and hence the reading accuracy could be expected to be in this range. Further, the orientation and placement of the device around the head are subject to manual error and this could impart in the detection performance. Automated compensation for orientation correction has not been discussed in this experiment and is a subject for future improvements. The work has been conducted on data from 28 human subjects and the model is theoretically robust with external data by usage of transformation techniques, however, this needs further validation by repeating the experiment on a wider range of human subjects. Further, the experiment has been conducted without contaminating the data from the same human for different sets which further validates the reliability of the model.

\subsection{Implementation scopes and practical usage of the work in real-world scenario}
The work could be used for improving Brain-Computer Interfacing. The EEG device may be attached to the helmets of riders where the system would monitor the moods and attention span of the riders and ultimately aid in avoiding accidents. The system could further be used in smart marketing recommendation systems for facilitating consumers with the exact product they need based on their mood. The mood of senseless patients could be monitored during surgeries to prevent patients from waking up during surgeries. The paralyzed or aphasia patients who have limited speaking ability could be monitored for their mood and aid could be provided accordingly.

\subsection{Limitations of the work}
The work has been conducted on 28 human subjects. This is a relatively much small number and therefore the proposed model needs to be validated on a much larger sample space. The work has been conducted to recognize 5 different moods and therefore the model further needs to be improved to recognize many more moods. The EEG signals sensitivity could be further increased to identify a much more faint but useful pattern.

\section{Conclusion}
\label{sec:conclusion}
As the brain is responsible for different moods, the associated EEG signals may be interpreted to detect those moods with a smart decision-making system developed by a neural network. The detection is further optimised by implementing Blackman window based Fourier Transformation and extracting the most significant frequencies for each electrode. 28 subjects with consent have been asked to play 4 games of different genres for 20 minutes each and their EEG readings have been recorded. Later these readings have been optimized with biomedical signal processing techniques and further convolution neural network algorithms have been used to detect the moods. For this purpose, 95.07\% detection accuracy has been obtained for detecting boring experiences followed by moods such as calm, horrible, funny and satisfaction with 94.89\%, 93.93\%, 93.57\% and 92.91\% accuracy respectively. The performances have been compared against a multi-layer perceptron and comparison between classification on raw data and optimized data have shown significant improvement in detection performance. Deployment of the model would be helpful to understand consumer behaviour, implement in road safety, etc. The model could further be developed for the detection of many more moods validating on a larger subject base.

\section*{Acknowledgement}
This experiment have been conducted as a segment of the Gyanam Project launched by a joint venture between Wingbotics and Spiraldevs Automation Industries Pvt. Ltd. West Bengal, India.

\bibliographystyle{ieeetr}

\bibliography{cas-dc-template}

\bio{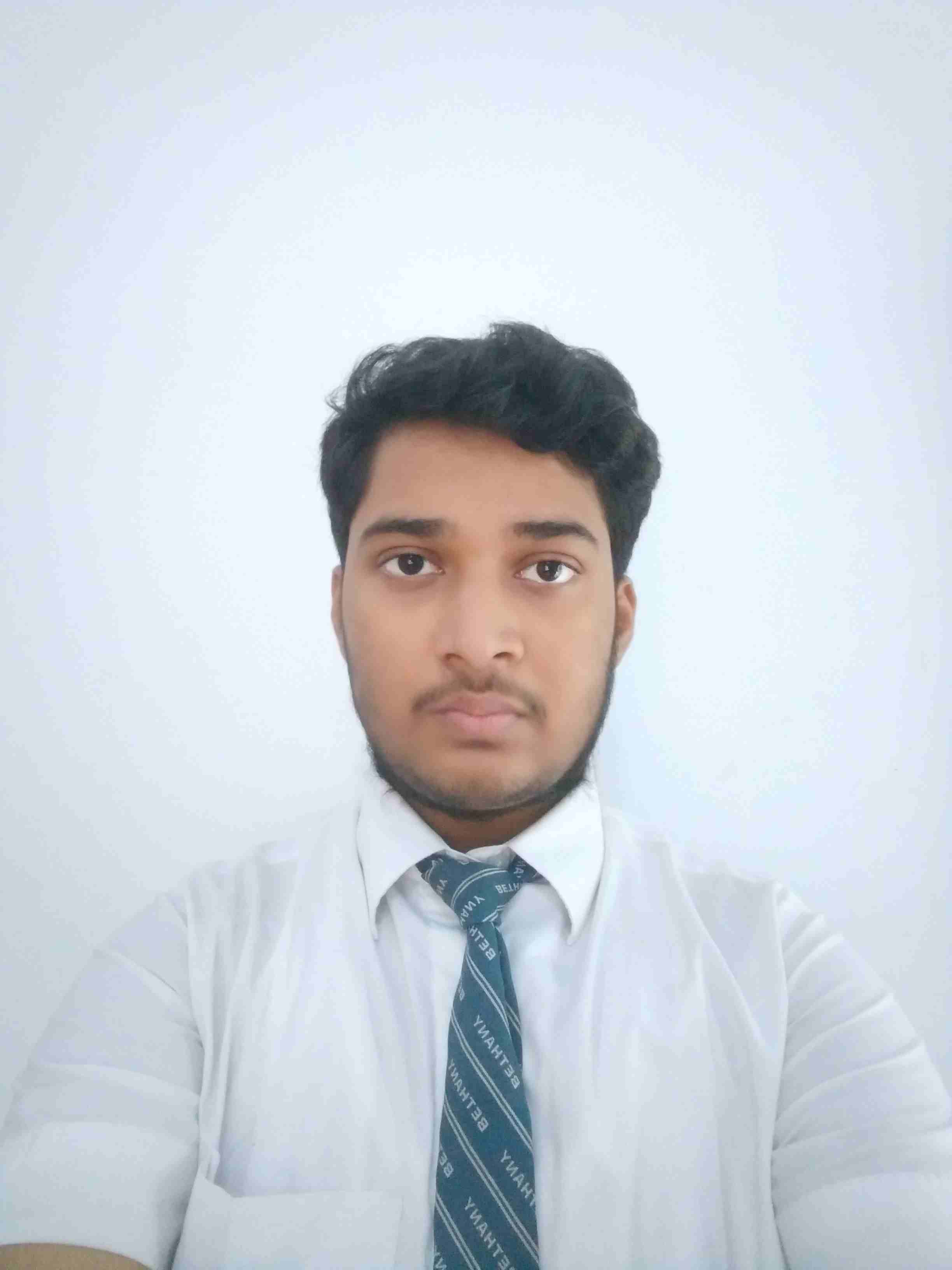}
\textbf{Subhrangshu Adhikary} is currently pursuing PhD from National Institute of Technology, Durgapur. He has completed his BTech on Computer Science and Engineering from Dr. B.C. Roy Engineering College, Durgapur, West Bengal, India during 2018-2022. He won the Best Paper Award in SMTST-2020 Conference, Uttarakhand, India powered by Springer. He has multiple research publications and patents. He have presented in top tier conferences like IGARSS and COMSNETS. He have also presented at 'Ocean Carbon from Space-2022' which is a workshop conducted jointly by ESA and NASA. He is the Director of Spiraldevs Automation Industries Pvt. Ltd., India which is an artificial intelligence and big data integrated full-stack development and IoT solutions startup. He is specialized in designing cost-efficient large-scale distributed systems capable of handling petabytes of data and billions of request-response per hour. He is also specialized in artificial intelligence focusing on machine learning and deep learning. Apart from this, he has expertise in multi-disciplinaries including cryptography and network security, cross-platform development, IoT, biomechanics, bioinformatics, bigdata, computer vision. He has designed several software systems which are continuously used by a wide range of industrial user base. He has also contributed to several peer reviews.
\endbio

\bio{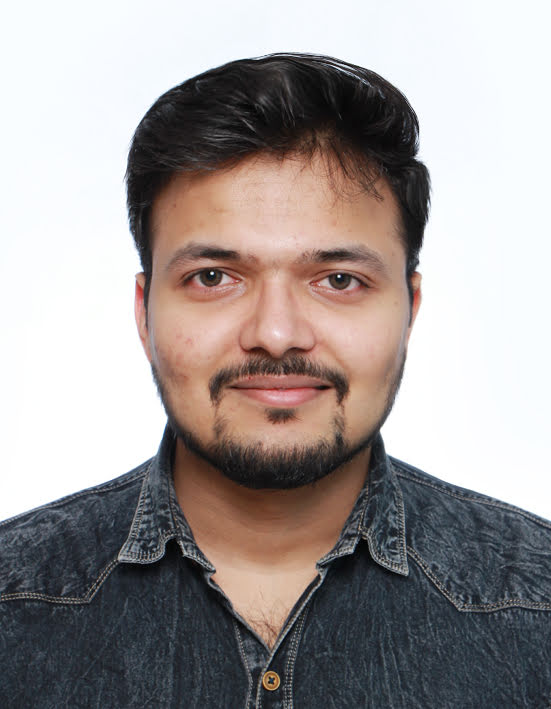}
\textbf{Kushal Jain} (MBBS) graduated in year 2020 from Harbin Medical University, China and served at Covid dedicated hospital Maulana Azad Medical College and Lok Nayak Hospital as Junior Resident Doctor. Currently He is working as Resident Doctor in department of Casualty at Vardhman Mahaveer Medical College and  Safdarjung Hospital, New Delhi. He has been actively involved in Research work and working on various projects
\endbio

\bio{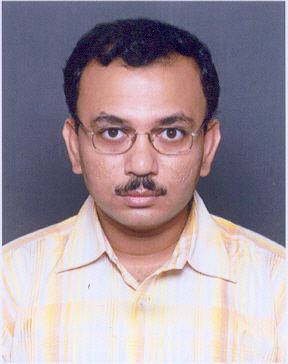}
\textbf{Biswajit Saha} was awarded Bachelor of Engineering in Computer Science and Engineering in the year 2002 and Master of Engineering in the year 2004. He is currently working as an Assistant Professor in an engineering college in India. His publications are on diverse topics such as Green Computing , Optimization and E Commerce etc. He has around nineteen years of teaching experience at the undergraduate and post graduate level of Engineering. His long term general and research interest is machine learning.
\endbio

\bio{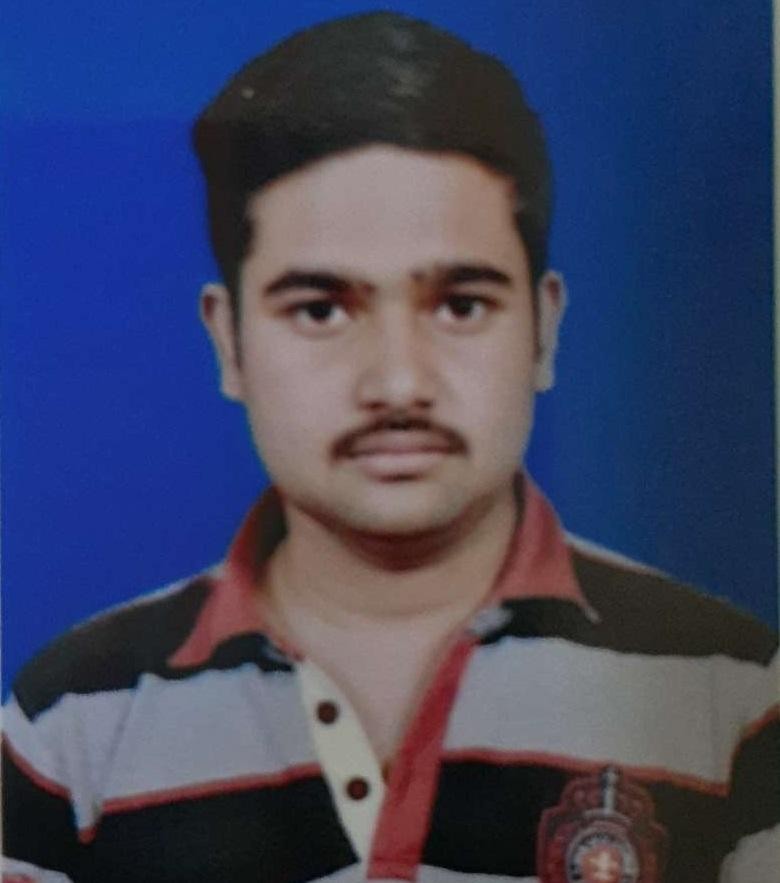}
\textbf{Deepraj Chowdhury} is with Department of Electronics and Communication Engineering, Dr. Shyama Prasad Mukherjee International Institute of Information Technology, Naya Raipur, India. He has Co-authored 10 research papers in different conferences like AusPDC 2022, CCGRID 2022, IEEE VTC 2022, INDICON 2021, ICACCP 2021 , IoT journal, MDPI Sensors, Wiley ITL. He also has 3 indian copyright registered, and applied for 3 Indian Patent. He is also serving as a reviewer for many reputed journals like Wiley Transaction on emerging Telecommunication Technologies, IEEE JBHI etc. His current research interests include IoT, Cloud Computing, Microwave, Cryptography, Blockchain, Artificial Intelligence.
\endbio
\end{document}